\documentstyle[aps,twocolumn,prl,epsf]{revtex}

\begin{document}

\newcommand{\be}{\begin{equation}}
\newcommand{\ee}{\end{equation}}

\draft

\twocolumn[\hsize\textwidth\columnwidth\hsize\csname @twocolumnfalse\endcsname

\title{Dielectric Response Near the Density-Driven Mott Transition
in infinite dimensions}
\author{Luis Craco$^1$ and Mukul S. Laad$^2$}

\address{$^1$Instituto de F\'{\i}sica ``Gleb Wataghin'' - UNICAMP, C.P. 6165, 
13083-970 Campinas - SP, Brazil \\ 
$^2$Institut fuer Theoretische Physik, Universitaet zu Koeln, Zuelpicher 
Strasse, 50937 Koeln, Germany  
}
\date{\today}
\maketitle

\widetext

\begin{abstract}
\noindent
We study the dielectric response of correlated systems which undergo a Mott 
transition as a function of band filling within the dynamical mean field 
framework. We compute the dielectric figure of merit (DFOM), which is a 
measure of dielectric efficiency and an important number for potential device 
applications.  It is suggested how the DFOM can be optimized in real 
transition metal oxides. The structures seen in the computed Faraday rotation 
are explained on the basis of the underlying local spectral density of the 
$d=\infty$ Hubbard model.
\end{abstract}


]

\narrowtext

The ac conductivity and dielectric tensor provides valuable information 
concerning the finite frequency, finite temperature charge dynamics of an
electronic fluid in a metal.  The potential for microwave devices, for e.g, 
is enormous, given the progress in synthesis of strongly correlated materials
with electronic properties that are sensitive to external perturbations like
applied electromagnetic fields, pressure(doping), etc~\cite{Imada}. Recent 
improvements in theoretical tools at our disposal allow controlled, 
physically meaningful calculations to be undertaken~\cite{Georges}. These 
advancements have, and will continue to, spur increased investigations to 
tap their full technological potential.

The dielectric figure of merit (DFOM) reflects the quality of the dielectric
response of a material, and is a quantity of interest for potential microwave
device applications.  It is formally defined as
\be
DFOM=i\frac{|\epsilon_{xy}(\omega)|}{2|Im \epsilon_{xx}(\omega)|}
\ee
and so one requires the full dielectric (ac conductivity) tensor to access this
quantity.  It also follows that it is linked to the magneto-optical response 
of the material under study.  In weakly correlated metals, the details of the
ac conductivity tensor, and hence of the dielectric and Hall response, are 
determined by the vagaries (shape and size) of the Fermi surface~\cite{Coleman}.  
That such a connection is untenable for strongly correlated metals has been 
pointed out by Shastry {\it et al}~\cite{Shas}, who show that the Hall 
constant, for e.g, is affected by contributions coming from the whole 

Brillouin zone, and may have nothing to do with Fermi surface effects.  
On the other hand, it has been observed that the physics of strongly 
correlated metals undergoing metal-insulator transitions is understandable 
in terms of spectral weight transfer over large energy scales~\cite{Georges}.
The consequences for dielectric response and the DFOM have, however, not 
been studied at all.

In this letter, we address this issue.  We are primarily interested in 
materials like $La_{1-x}Y_{x}TiO_{3}$~\cite{Imada}, which undergo Mott 
transitions with doping.  We stress that 3D transition metal oxides are the 
most interesting candidates for the kind of effects we want to study, as 
filling driven insulator-metal transitions are realized in a variety of 
them in a wide range of parameters.

We consider the one-band Hubbard model~\cite{Hubbard},
\be
\label{model}
H=-\sum_{ij\sigma}t_{ij}c_{i\sigma}^{\dag}c_{j\sigma} 
+ U\sum_{i}n_{i\uparrow}n_{i\downarrow} - \mu\sum_{i}n_{i\sigma}
\ee
as a prototype model describing the electronic degrees of freedom in TM 
oxides. 
To study the 3D case, we employ the $d=\infty$ approximation, which is the 
best approximation possible at the present time~\cite{Georges}.
Since this method has been extensively reviewed, we only 
summarize the relevant aspects. All transport properties, which follow from 
the conductivity tensor, are obtained from a ${\bf k}$-independent 
self-energy in $d=\infty$; the only information about the lattice structure 
comes from the free band dispersion in the full Green fn:
\be
G(k,\omega)=G(\epsilon_{k},\omega)=\frac{1}
{\omega+\mu-\epsilon_{k}-\Sigma(\omega)}
\ee
To solve the model in $d=\infty$ requires a reliable way to solve the single 
impurity Anderson model(SIAM) embedded in a dynamical bath described by the
hybridization fn. $\Delta(\omega)$.  There is an additional condition that 
completes the selfconsistency:
\be
\int d\epsilon G(\epsilon, i\omega)\rho_{0}(\epsilon) = \frac{1}
{i\omega+\mu-\Delta(i\omega)-\Sigma(i\omega)}
\ee
where $\rho_{0}(\epsilon)$ is the free DOS ($U=0$).  In $d=\infty$, this is 
sufficient to compute the transport, because the vertex corrections in the 
Bethe Salpeter eqn. for the conductivity vanish identically in this 
limit~\cite{Khurana}. Thus, the conductivity is fully determined by the basic 
bubble diagram made up of fully interating local GFs of the lattice model.

The optical conductivity and the Hall conductivity are computable in terms 
of the full $d=\infty$ GFs as follows~\cite{Georges}:
\be
\sigma_{xx}(i\omega)=\frac{1}{i\omega}\int 
d\epsilon\rho_{0}(\epsilon)\sum_{i\nu}G(\epsilon,i\nu)G(\epsilon,i\nu+i\omega)
\ee
and the Hall conductivity has been worked out by Lange~\cite{Lange}, so we use 
the approach developed there.  Explicitly, after a somewhat tedious 
calculation, the imaginary part of $\sigma_{xy}(\omega)$ is given by

\begin{eqnarray}
\nonumber
\sigma_{xy}^{''}(\omega) &= & c_{xy}\int_{-\infty}^{+\infty} d\epsilon
\rho_{0}(\epsilon)\epsilon \int_{-\infty}^{+\infty} d\omega_{1}
d\omega_{2} A(\epsilon,\omega_{1})A(\epsilon,\omega_{2})
\\ &&\frac{1}{\omega} \left[
\frac{F(\epsilon,\omega_{1};\omega)-
F(\epsilon,\omega_{2};\omega)}{\omega_{1}-\omega_{2}}
+(\omega \rightarrow -\omega) \right]
\end{eqnarray}
where

\be
F(\epsilon, \omega; \omega_{1}) = A(\epsilon, \omega_{1}-\omega)
[f(\omega_{1})-f(\omega_{1}-\omega)]
\ee
and $A(\epsilon,\omega)=-Im [\omega-\epsilon-\Sigma(\omega)]^{-1}/\pi$ is the
s.p spectral function in $d=\infty$.  

Knowledge of $\sigma_{xy}^{''}(\omega)$ permits us to use the analyticity 
properties to compute its real part via a Kramers-Kr\"onig transform.  The
dielectric tensor is determined from,
\be
\epsilon_{xx}(\omega)=1+\frac{4\pi}{\omega}i\sigma_{xx}(\omega)
\ee
and
\be
\epsilon_{xy}(\omega)=\frac{4\pi}{\omega}i\sigma_{xy}(\omega)
\ee
The paramagnetic Faraday rotation is directly related to  $\sigma_{xy}(\omega)$
via~\cite{Shas}
\be
\label{tf}
\theta_{F}(\omega)=C(n)i\frac{\sigma_{xy}^{''}(\omega)}{\omega}
\ee
The DFOM can readily be computed from the dielectric tensor, as determined 
above.  So can the ac Hall constant and angle, as well as the Raman intensity,
which has been discussed in detail elsewhere~\cite{Freer,Laad}.  

Before embarking on our results and their analysis, it is instructive to 
summarize what is known about the $d=\infty$ Hubbard model.  At large $U/t$, 
and away from half-filling ($n=1$), the ground state is a paramagnetic FL if 
one ignores the possibility of symmetry breaking towards antiferromagnetism,
as well as disorder effects, which are especially important near $n=1$.  This
can be achieved formally by introducing a nnn hopping, which in $d=\infty$
leaves the free DOS essentially unchanged~\cite{Georges}.  This metallic state 
is characterized by two energy scales: a low energy coherence scale $T_{coh}$, 
below which local FL behavior sets in~\cite{Georges}, and a scale of $O(D)$, 
($D$ is the free bandwidth) characterizing high energy, incoherent processes 
across the remnant of the Mott-Hubbard insulator at $n=1$.  At $T<T_{coh}$, 
the quenching of the local moments leads to a response characteristic of a FL 
at small $\omega<<2D$ (but with the dynamical spectral weight transfer with 
doping, a feature of correlations) , but at higher $T>T_{coh}$, the picture 
is that of carriers scattered off by effectively local moments, which makes 
the system essentially like a non-FL (note that the metal with disordered 
local moments is not a FL).

\begin{figure}[htb]
\epsfxsize=3.5in
\epsffile{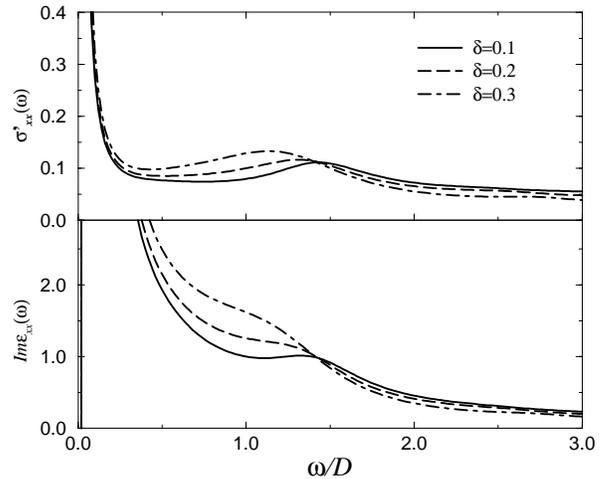}
\caption{Real part of the optical conductivity $\sigma_{xx}$ for $U/D=3.0$, 
$\delta=0.1$ (continuous), $\delta=0.2$ (dashed) and $\delta=0.3$ 
(dot-dashed). The lower panel shows the imaginary dielectric constant for
the same parameter values.  Note the isosbectic point in both quantities.}   
\label{fig1}
\end{figure}

Armed with this information, we are ready to discuss our results.
We choose a gaussian unperturbed DOS, and $U/D=3.0$ to access the strongly
correlated FL metallic state off $n=1$, ignoring the possible instability to
an AF-ordered phase.  All calculations are performed at a low temperature,
$T=0.01D$. Fig.~\ref{fig1} shows the optical conductivity and the longitudinal 
dielectric constant.  $\sigma_{xx}(\omega)$ agrees with calculations 
performed earlier in all the main respects; in particular, it clearly 
exhibits the low-energy quasicoherent Drude form, the transfer of optical 
spectral weight from the high-energy, upper-Hubbard band states to the low 
energy, band-like states with increasing hole doping, and the isosbectic 
point at which {\it all} the $\sigma_{xx}(\omega)$ curves as a fn. of filling 
cross at one point, to within numerical accuracy.  It is interesting to point 
out that such features have also been observed in experimental 
studies~\cite{Imada,Uchida}. Correspondingly, 
Im$\epsilon_{xx}(\omega)$ also shows 
the isosbectic point, the explanation for which is identical to that provided 
recently by us for the case of $\sigma_{xx}(\omega)$~\cite{Laad}.

\begin{figure}[htb]
\epsfxsize=3.5in
\epsffile{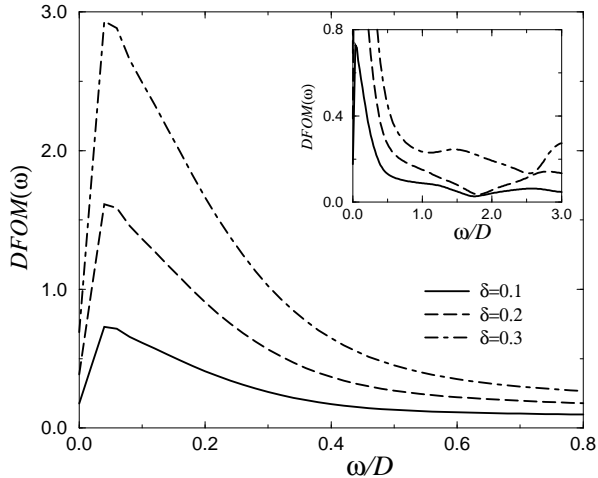}
\caption{  The dielectric figure of merit (DFOM) for the doped Hubbard model 
for $U/D=3.0$, and $\delta=0.1, 0.2, 0.3$ as shown.  See text for more 
details.}
\label{fig2}
\end{figure}

In Fig.~\ref{fig2}, we show the DFOM obtained using eqn.~(\ref{model}) and the 
dielectric tensor calculated as above.  We are mainly interested in the 
variation of the DFOM with hole doping, given here by $\delta=(1-n)$.  This 
fixes the chemical potential, and the FL resonance position, and the IPT 
describes the evolution of spectral features in good agreement with exact 
diagonalization studies~\cite{Georges}. In view of the ability of the IPT to 
reproduce all the qualitative aspects observed in 
$\sigma_{xx}(\omega)$, we believe that is a good tool in the 
present case.

The DFOM shows a sharp peak at low energy (around $0.05$ ev) in the metallic 
state off half-filling with a maximum value of order 3 for $\delta=0.3$.  It 
also reaches a minimum value at around 1.8 ev for $\delta=0.1, 0.2$; this is
related to the frequency dependence of the optical conductivity tensor with 
filling.  These features are understandable in terms of the dynamical transfer
of optical spectral weight from the high-energy (upper-Hubbard band) incoherent
states to low energy quasicoherent states upon hole doping the Mott insulator
(at $n=1$).  The sharp peak is related to the fact that the action of the 
current operator on the lower-Hubbard band states creates well defined 
elementary excitations in a strongly correlated Fermi liquid, while the 
increase of the 
low energy weight is understood in terms of the increasing weight of the 
quasicoherent, itinerant part of the spectrum relative to that of the atomic
like, incoherent part with increasing $\delta=(1-n)$.  Thus, the absolute DFOM
is determined by the outcome of the competition between the itinerant and 
atomic parts of the spectrum.  The above suggests that the DFOM in the IR and
the mid-IR can be increased even further by enhancing the weight of the 
transitions within the lower-Hubbard band and those in the (lower-Hubbard
band + central FL peak) manifold, by suitable materials engineering.  In 
reality, a multi-orbital situation would be more favorable, since one might
expect increased contributions to the mid-IR part coming from interorbital
hopping, as well as from spin-orbit coupling terms present in multi-orbital
systems.  However, static disorder, for e.g, in $d=\infty$, would shift 
coherent spectral weight to higher energy and destroy low energy coherence,
limiting the DFOM to modest values, suggesting that good sample quality is one
of the prerequisites to increase the DFOM (something that maybe hard to limit 
in the 3d transition metal compounds undergoing doping/filling driven Mott
transitions).

\begin{figure}[htb]
\epsfxsize=3.5in
\epsffile{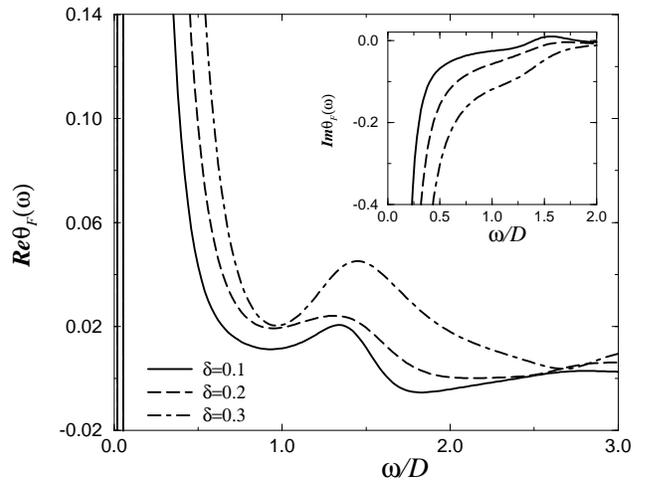}
\caption{  The paramagnetic Faraday rotation $\theta_{F}(\omega)$ for the same
parameter values as in Fig.~\ref{fig1}. See the text for the explanation of 
various structures in $\theta_{F}(\omega)$ in the light of the underlying 
structure of the local spectral density of the HM in $d=\infty$.}
\label{fig3}
\end{figure}

Next, we describe our results for the paramagnetic Faraday rotation, computed
from eqn.~(\ref{tf}).  $\theta_{F}(\omega)$ shows a monotonic fall-off with 
$\omega$ in the IR and the mid-IR region, before increasing around 
$\omega/D \simeq 1$,
whereafter, it passes through a broad maximum at $\omega/D \simeq 1.4$.  
This is followed by another broad peak around $\omega/D = 3.0$.  The origin 
of these features are linked to the nature of the transitions reflected in 
the interacting LDOS of the $d=\infty$ Hubbard model.  We assign the central 
feature to the ``quasipartcle peak'' near $\omega = 0$, the maximum feature 
around $\omega/D \simeq 1.4$ to the ``$U/2$ peak'', corresponding to 
transitions between the QP peak and the upper- (or lower) Hubbard bands, 
and the broad feature at $\omega/D = 3.0$ to transitions between the 
lower- and the upper Hubbard bands.

The above suggests that strongly correlated metals near the borderline of the
filling-driven Mott transition might be good candidates for dielectric device
applications.  A multi-orbital situation (plus spin-orbit couplings) would be
more effective in enhancing the DFOM, while static disorder, which inevitably 
accompanies doping, would act to limit it.  Consideration of real bandstructure
would be desirable; in our approach, this simply entails replacing the free
(gaussian) DOS used here by the actual bandstructure DOS as an input into the
DMFT calculation.  Effects of static disorder will be especially important 
near the doping induced Mott transition; these require the consideration of 
correlations and disorder on an equal footing, and is left for a future work.
However, one expects that the destruction of low energy coherent spectral 
weight which accompanies static disorder will limit the DFOM to more modest 
values.

  In conclusion, we have investigated the dielectric figure of merit (DFOM) and
the paramagnetic Faraday rotation near the density driven Mott transition.  We
have shown how a consistent treatment of the interplay between the atomic and 
itinerant aspects inherent in strongly correlated systems leads to encouraging
values for the DFOM.  In practice, however, the estimate provided here would
be an overestimate, since real bandstructure and disorder effects, as well as
multi-orbital character of real $d$-band systems can well change our 
conclusions quantitatively.  We have also provided a simple explanation for 
the origin of structure in the Faraday rotation; in real materials, the above 
effects will also affect $\theta_{F}(\omega)$.  Nevertheless, we have 
provided a theoretical framework within which such calculations can be 
undertaken, and all of the above effects can be included in a suitable 
extension within the $d=\infty$ framework.

\acknowledgments
One of us (MSL) thanks Prof E. M\"uller-Hartmann for useful 
discussions, and SFB341 for financial support.  LC acknowledges financial
support of Funda\c c\~ao de Amparo \`a Pesquisa do Estado de 
S\~ao Paulo (FAPESP).


\begin{references}

\bibitem{Imada} M. Imada, {\it et al.}, Revs. Mod. Phys. {\bf 70}, 1039 
(1998). 

\bibitem{Georges} A. Georges {\it et al.}, Revs. Mod. Phys. {\bf 68}, 13 
(1996). 
\bibitem{Coleman} P. Colemam {\it et al.}, J. Phys. Cond. Mat. {\bf 8}, 9985 
(1996).                                

\bibitem{Shas} B. S. Shastry {\it et al.}, Phys. 
Rev. Lett. {\bf 70}, 2004 (1993); {\bf 71}, 2838 (1993).

\bibitem{Hubbard} J. Hubbard, Proc. Roy. Soc. London, Ser. A{\bf 276}, 238 
(1963). 

\bibitem{Khurana} A. Khurana, Phys. Rev. Lett. {\bf 64}, 1990 (1990).

\bibitem{Lange} E. Lange, Phys. Rev. B {\bf 55}, 3907 (1997).

\bibitem{Freer} M. Jarrell {\it et al.}, Phys. Rev. B{\bf 51}, 11704 
(1995).

\bibitem{Laad} M. S. Laad {\it et al.}, cond-mat 9907328. 

\bibitem{Uchida} A great variety of transition metal and rare-earth compounds
exhibit the interesting phenomenon of isosbectic points in the 
$\sigma_{xx}(\omega)$ vs $\omega$ curves.  See Ref.~\cite{Imada} for many 
systems off half-filling where this feature is observed experimentally. 

\end{references}
\end{document}